\documentclass[namedreferences]{SolarPhysics}
\usepackage[optionalrh]{spr-sola-addons} 
\usepackage{graphicx}        
\usepackage{color}           
\usepackage{url}             

\newcommand{\eqref}[1]{(\ref{eqn:#1})}
\newcommand{\expt}[1]{\,\textrm{e}^{#1}}

\newcommand{\megam}{\,\textrm{Mm}}

\newcommand{\mpsec}{\,\textrm{m}\,\textrm{s}^{-1}}
\newcommand{\kmpsec}{\,\textrm{km}\,\textrm{s}^{-1}}
\newcommand{\kmsqpsec}{\,\textrm{km}^2\textrm{s}^{-1}}

\newcommand{\etal}{{\it et al.}}

\renewcommand{\vec}[1]{ {\mathbf #1} }

\newcommand{\pder}[2]{ \frac{\partial #1}{\partial #2} }

\newcommand{\avec}{ \vec A}

\newcommand{\bb}{ \vec B}
\newcommand{\jj}{ \vec j}

\begin{document}

\begin{article}

\begin{opening}

\title{Modelling the Global Solar Corona II: Coronal Evolution and Filament Chirality Comparison}

\author{A.R.~\surname{Yeates}$^{1}$\sep
        D.H.~\surname{Mackay}$^{1}$\sep
        A.A.~\surname{van Ballegooijen}$^{2}$      
       }
\runningauthor{A.R.~Yeates \etal}
\runningtitle{Modelling the Global Solar Corona II}

   \institute{$^{1}$ School of Mathematics and Statistics, University of St Andrews, Fife, KY16 9SS, Scotland
                     email: \url{anthony@mcs.st-and.ac.uk} \\ 
              $^{2}$ Harvard-Smithsonian Center for Astrophysics, 60 Garden Street, Cambridge, MA 02138, USA \\
             }

\begin{abstract}
 The hemispheric pattern of solar filaments is considered using newly-developed simulations of the real photospheric and 3D coronal magnetic fields over a 6-month period, on a global scale. The magnetic field direction in the simulation is compared directly with the chirality of observed filaments, at their observed locations. In our model the coronal field evolves through a continuous sequence of nonlinear force-free equilibria, in response to the changing photospheric boundary conditions and the emergence of new magnetic flux. In total 119 magnetic bipoles with properties matching observed active regions are inserted. These bipoles emerge twisted and inject magnetic helicity into the solar atmosphere. When we choose the sign of this active-region helicity to match that observed in each hemisphere, the model produces the correct chirality for up to 96\% of filaments, including exceptions to the hemispheric pattern. If the emerging bipoles have zero helicity, or helicity of the opposite sign, then this percentage is much reduced. In addition, the simulation produces a higher proportion of filaments with the correct chirality after longer times. This indicates that a key element in the evolution of the coronal field is its long-term memory, and the build-up and transport of helicity from low to high latitudes over many months. It highlights the importance of continuous evolution of the coronal field, rather than independent extrapolations at different times. This has significant consequences for future modelling such as that related to the origin and development of coronal mass ejections.
\end{abstract}
\end{opening}

\section{Introduction}

The recent work of \inlinecite{mackay3} gives a convincing explanation for the hemispheric pattern of filament chirality on the Sun. A filament may be classified as having either dextral or sinistral chirality (handedness) depending on whether its axial magnetic field points right or left, as seen from the side of the filament with positive magnetic polarity in the photosphere. Observations show that quiescent filaments in the northern hemisphere tend to be dextral, while those in the southern hemisphere tend to be sinistral \cite{rust1967,leroy1983,martin1994,pevtsov2003}. The latter paper found that $80\%$ -- $85\%$ of quiescent filaments follow this hemispheric pattern, which remains the same when the Sun's polar field reverses every 11 years.

Using the model of \inlinecite{vanballegooijen2000} to simulate the coupled evolution of photospheric and coronal magnetic fields, \inlinecite{mackay3} have shown that the dominant chirality may be produced for the dominant range of bipole tilt angles \cite{wang1989} and dominant sign of helicity \cite{pevtsov1995} observed in each hemisphere. The model suggests how the evolution of bipolar active regions, and their interaction as they are transported poleward, may explain both the sheared magnetic fields in filament channels and their hemispheric pattern. Importantly, the authors showed how exceptions to the hemispheric pattern could occur in a natural way, for bipoles with very large tilt angles or the minority sign of helicity for their hemisphere. However, as in the filament-formation simulations of \inlinecite{litvinenko2005} and \inlinecite{welsch2005}, \inlinecite{mackay3} considered only a pair of magnetic bipoles, in a localised region of the solar corona. In this paper we test their conclusions in a direct comparison between global coronal simulations and observations of many real filaments on the Sun, over a 6-month period in 1999.

In the first paper (\opencite{yeates2007a}, hereafter ``Paper 1''), we described how barb orientations were used to determine the chirality of 255 filaments observed in H$\alpha$ images from Big Bear Solar Observatory (BBSO). In the present paper, the skew of the simulated magnetic field is compared with the chirality of these filaments, at their observed locations. This comparison requires the extension of the previous simulations to cover the global corona, and its evolution over a period of many months.

In Paper 1, we described the photospheric simulations, which provide a continuously evolving lower boundary condition for the 3D coronal field. By inserting new active regions into the simulation as they were observed, we showed how accuracy to the observed synoptic magnetograms could be maintained without ever resetting the simulation to the observed photospheric field. In particular large areas of positive and negative flux were well reproduced, as were the shape of the polarity inversion lines (PILs) at the majority of observed filament locations. This represents a significant improvement over previous simulations of filament formation in the real corona \cite{mackay2}, which lost accuracy over time if the photospheric field was not reset to the observed magnetogram every 27 days. The solution they adopted was to construct a new potential field in the corona at this point and begin the evolution again, thus removing any long-term memory in the coronal field of previous interactions, as well as any helicity that had built up. Magnetic helicity is a quantity that measures the topological properties of field lines such as linking, twist, or kinking (\opencite{berger1998}, \citeyear{berger1999}). We believe that the gradual build-up of helicity, in the form of sheared non-potential field lines, is key to the hemispheric pattern of filaments. Our new simulations are able to maintain accuracy in the photospheric field over many months without resetting, enabling a continuous evolution of the coronal field and so no loss of memory.

Potential magnetic fields, which contain no electric currents, are the simplest way to model the magnetic field in the solar corona. Extrapolations using an upper ``source-surface'' where the field is assumed to be radial were first proposed in the 1960s \cite{altschuler1969,schatten1969}, and have since been used by many authors to study the evolution of the coronal field based on photospheric motions (\textit{e.g.} \opencite{wang2002}; \opencite{mackay5b}; \opencite{lockwood2003}; \opencite{schrijver2003}). However, in these models, the coronal field is extrapolated from the photospheric simulations at discrete intervals. It is independently computed at each time and contains no memory of its past evolution. The potential field model does not describe \emph{how} the magnetic field of the Sun evolved into a particular configuration. It precludes the long-term build-up of magnetic helicity and shear which have been shown to be essential for filament formation \cite{mackay2003}, and in particular to model the observed hemispheric pattern \cite{mackay3}. Indeed potential fields represent the state of minimum magnetic energy, so cannot represent the sheared fields of active regions \cite{riley2006, regnier2007} or filament channels \cite{mackay1997}. 

A more realistic model for such regions is a force-free equilibrium, where the Lorentz force $\vec{j}\times\bb$ vanishes everywhere but the current density $\vec{j}$ may be non-zero. This approximation is thought to be good for magnetic fields in the corona \cite{priest1982}, where the magnetic pressure is generally much greater than the gas pressure, so that other forces may be neglected and the Lorentz force must balance itself. This condition may be written
\begin{equation}
\nabla\times\bb = \alpha\bb
\label{eqn:forcefree}
\end{equation}
for some $\alpha$. If $\alpha$ is everywhere constant, this is called a linear force-free field, and Equation \eqref{forcefree} is readily solved. The special case $\alpha=0$ is the potential field discussed earlier, while constant non-zero values have been used successfully to model individual filaments \cite{aulanier1998,aulanier2000}. However, such constant-$\alpha$ fields are limited to localised regions of the corona, and in general $\alpha$ can be a function of position, although it must be constant along any individual field line. The more realistic fields where $\alpha$ is a function of position are called nonlinear force-free, and are considerably more difficult to compute, particularly with regard to the required boundary conditions. Various methods have been developed to compute nonlinear force-free fields based on a single surface vector-magnetogram (see \opencite{schrijver2006} for a summary), and are being developed to compute the global coronal field based on a full-disk vector-magnetogram \cite{wiegelmann2007}.

The technique we adopt here is described in Section \ref{sec:model} and differs significantly from these extrapolation methods. It effectively produces a continuous sequence of nonlinear force-free equilibria as the 3D coronal field responds to flux emergence and photospheric motions, rather than a single-time extrapolation. In this paper we consider whether our model can produce the correct hemispheric pattern observed in our filament sample. Section \ref{sec:skew} presents the simulation results; firstly we consider the overall percentage of simulated filaments following the hemispheric pattern (Section \ref{sec:pattern}), and then we compare the chirality of individual filaments with the observations (Section \ref{sec:indiv}). The performance of the simulations is discussed in Section \ref{sec:discussion}. Conclusions are given in Section \ref{sec:conclusion}, along with consequences for future modelling.

\section{Coronal Field Model} \label{sec:model}

\begin{figure}
\centerline{
\includegraphics[width=0.475\textwidth,clip=]{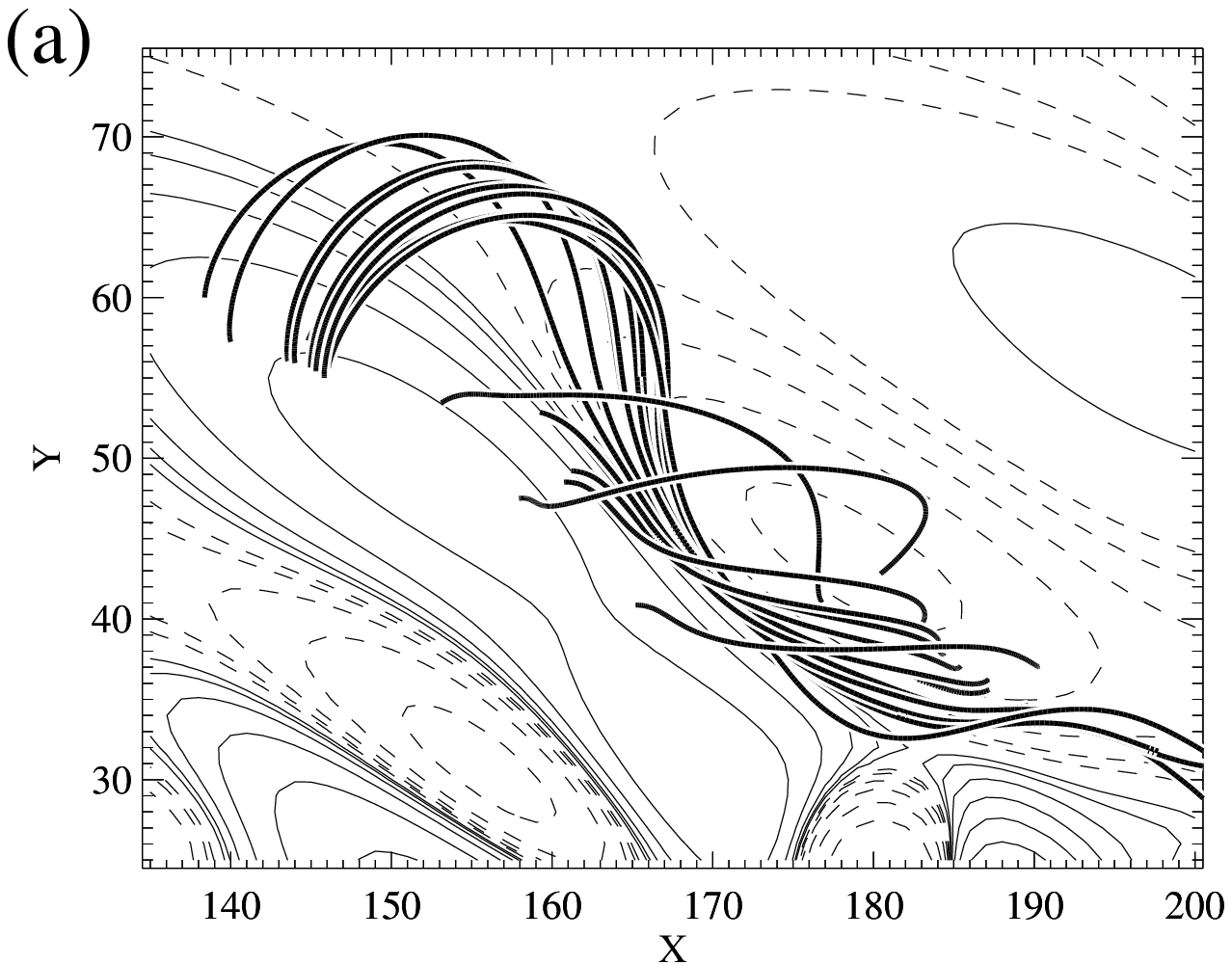}
\includegraphics[width=0.45\textwidth,clip=]{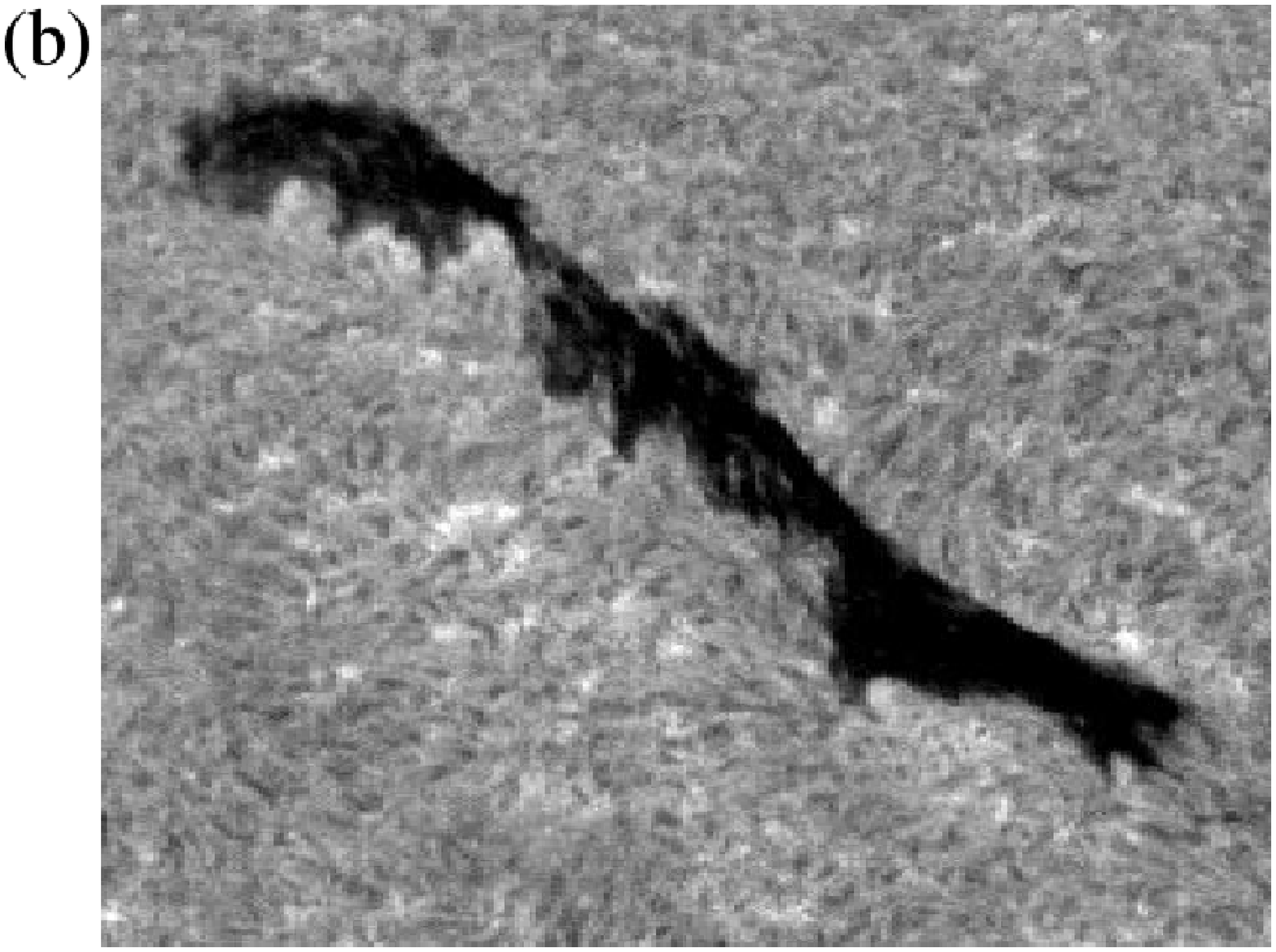}
}
\centerline{
\includegraphics[width=0.9\textwidth,clip=]{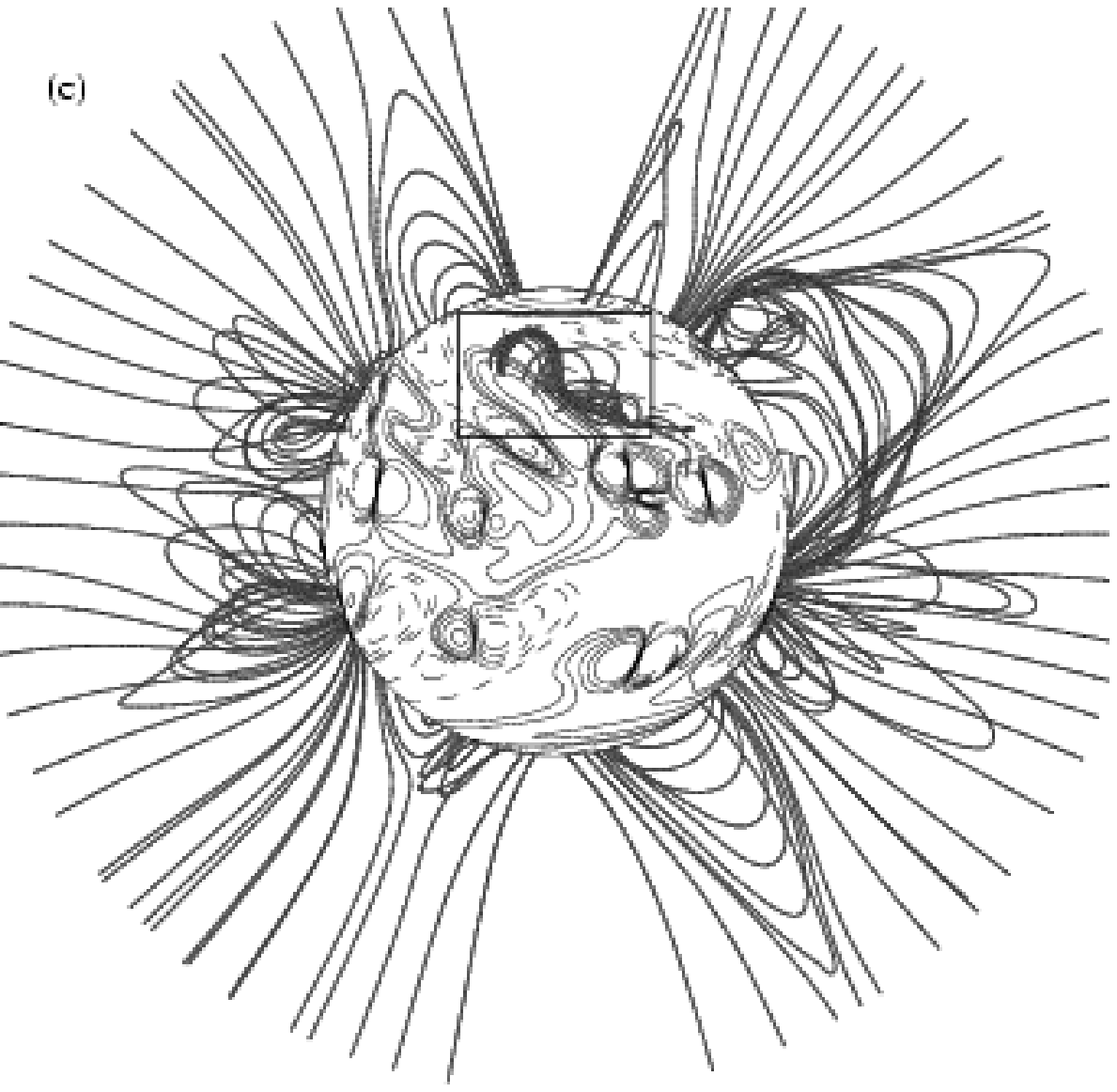}
}
\caption{Comparison for filament 544: (a) simulated flux rope structure (with bipole twist $\left|\beta\right|=0.2$, correct hemispheric sign), (b) BBSO H$\alpha$ image of the real filament on 14 September 1999, and (c) context of the filament in the global simulation. In (a) and (c), radial magnetic field strength on the photospheric surface is shown by contours (solid for positive flux and dashed for negative flux), while thicker lines denote selected coronal field lines. The box on panel (c) gives the approximate location of panels (a) and (b).}
  \label{fig:filament544}
\end{figure}

In this section we describe our model of the 3D coronal field, which is based on that of \inlinecite{vanballegooijen2000}. The model has two basic components:
\begin{enumerate}
\item
At the lower boundary of the 3D model, the radial magnetic field in the photosphere evolves under the effects of (1) the emergence of new magnetic flux from below the photosphere, (2) the advection of flux by the large-scale motions of differential rotation and meridional flow, (3) the dispersal of flux by small-scale convective cells, and (4) the cancellation of flux at polarity inversion lines (PILs). These are all observed effects (see {\it e.g.} \opencite{wang1989b}). The last two processes are described by including a diffusion term in the flux-transport equations, which is a standard technique that has been used in solar physics for many years \cite{sheeley2005}.
\item
The corona evolves in response to flux emergence and the changing photospheric boundary conditions: (1) New magnetic bipoles emerge into the corona, displacing older fields and producing currents at the interface between the old and new flux systems. (2) The new fields are assumed to be twisted, producing yet other currents and causing helicity to be injected into the atmosphere. (3) The horizontal motions at the lower boundary cause shearing of the coronal field, and surface diffusion causes reconnection of coronal field lines, which modifies the coronal currents. (4) After some period of time the magnetic field becomes unstable, and eruptions occur that cause twisted fields to be ejected into the heliosphere \cite{mackay4}. To describe these processes, we solve a reduced form of the MHD equations. This form assumes that the coronal Alfv\'en speed is high, and that the coronal magnetic field reacts quickly to the changing photospheric boundary conditions. In essence, the corona evolves through a series of force-free equilibria \cite{antiochos1987}.
\end{enumerate}
Figure \ref{fig:filament544}(c) shows a snapshot of the global photospheric and coronal fields in the simulation. Selected coronal field lines have been chosen for clarity, and illustrate the twisted, non-potential nature of the 3D field.

The surface flux transport model, which provides the lower boundary condition in our 3D simulations, has been described in Paper 1. In this paper we summarise the observational inputs to the model (Section \ref{sec:observational}), before outlining the reduced MHD equations for the coronal field (Section \ref{sec:coronal}) used in our numerical model (Section \ref{sec:numerical}). Our new method for incorporating emerging active regions was described in detail in Paper 1. In Section \ref{sec:emerging} of this paper we consider the insertion of these regions in the 3D corona, and we discuss some important parameters of the regions in Section \ref{sec:parameters}.

\subsection{Observational Inputs} \label{sec:observational}

The ideal input to the simulation would be measurements of the vector magnetic field $\bb$ of all emerging bipoles, both on the solar surface and in the corona above. Unfortunately the magnetic field cannot be directly observed in the corona, and the only component of $\bb$ readily available on the surface is the radial component $B_r$. Vector magnetograms giving all three components exist only for very localised regions on the solar surface, and the use of these observations to construct coronal fields is a difficult computational problem \cite{schrijver2006} so not useful for our study. To consider the coronal field evolution on a time-scale of months, we must therefore carry out modelling based on the radial magnetic field on the solar surface.

We use synoptic maps of $B_r$ constructed by the U.S. National Solar Observatory, Kitt Peak. Each map is essentially a representation of the radial magnetic field at the Sun's central meridian as a function of time over one Carrington rotation. In Paper 1 we described how these maps are used in two ways as input to our surface simulations. Firstly an initial condition for the simulation is produced by correcting a synoptic map for the effects of differential rotation, so that it best represents a single day. Secondly, the locations of newly-emerged magnetic flux throughout the simulation are determined by comparing subsequent synoptic maps, which are again corrected for differential rotation. The resulting surface field contains the observational input to the 3D coronal model described in this paper, which evolves in response to the flux emergence and changing photospheric boundary conditions.

\subsection{Coronal Evolution Equations} \label{sec:coronal}

In the corona we apply the technique used in \inlinecite{mackay4}. The magnetic field evolves in response to the non-ideal induction equation. As we solve for the vector potential $\avec$, where $\bb = \nabla \times \avec$, the equation takes the form
\begin{equation}
\pder{\avec}{t} = \mathbf{v}\times\bb - \eta_c\,\jj,
\label{eqn:induction}
\end{equation}
where $\mathbf{v}(\mathbf{r},t)$ is the plasma velocity, $\eta_c$ is the coronal diffusivity, and $\jj=\nabla\times\bb$. The gauge has been chosen so that there is no additional gradient term on the right-hand side of this equation.

Rather than solving the full momentum equation, we follow the magnetofrictional method of \inlinecite{yang1986} and assume the coronal plasma velocity to take the form
\begin{equation}
\mathbf{v} = \frac{1}{\nu}\frac{\jj\times\bb}{|\bb|^2} + v_0\expt{-\left(2.5R_\odot-r\right)/r_w}\mathbf{\hat{r}}.
\label{eqn:velocity}
\end{equation}
Of course, this reduced model does not include plasma properties such as density or temperature, but the magnetic field is known to dominate the plasma in the corona, and our technique is much less computationally intensive than full-MHD models of the corona (\textit{e.g.} \opencite{mikic1999}). These would not be practical for this study where we consider the evolution over a period of months.

The first term on the right-hand side of Equation \eqref{velocity} represents in an approximate manner the dominance of the Lorentz force in the corona. It corresponds to a frictional force causing the plasma to relax toward a force-free state. The second term in Equation \eqref{velocity} is a radial outflow velocity imposed to ensure the field lines remain radial at the outer boundary $r=2.5R_\odot$ of the computational domain, and simulating in a crude fashion the effect of the solar wind opening up coronal field lines. This term has a peak value $v_0=100\kmpsec$ and falls off exponentially away from the top boundary, so that there is only a small increase in the open magnetic flux compared to a potential field, and not all field lines are blown open. As in the previous simulations of \inlinecite{mackay4}, there are localised temporary losses of equilibrium as flux ropes are ejected from the computational box. These are not critical for the present study but will be considered in more detail in a subsequent paper.

The coronal diffusivity is taken to have the form
\begin{equation}
\eta_c = \eta_0\left(1 + 0.2\frac{|\jj|}{|\bb|}\right),
\label{eqn:diffusivity}
\end{equation}
as in \inlinecite{mackay4}. This diffusion was introduced to limit the twist in helical flux ropes to less than one turn, as is usually observed in filaments, because the previous ideal-MHD simulations \cite{mackay2001} produced highly-twisted structures. It includes both a background diffusivity of strength $\eta_0 = 0.1D$, where $D=450\kmsqpsec$ is the photospheric diffusivity (Paper 1), and an enhanced term which acts only in regions of strong current density.

\subsection{Numerical Method} \label{sec:numerical}

Equation \eqref{induction} is solved in a three-dimensional domain representing the solar corona in both hemispheres, where $R_\odot \leq r \leq 2.5R_\odot$, $10^\circ \leq \theta \leq 170^\circ$, and $0^\circ \leq \phi \leq 360^\circ$. The computations are carried out in transformed coordinates $(x,y,z)$, where
\begin{equation}
x=\phi/\Delta, \quad y = -\ln\left(\tan\left( \theta/2\right) \right)/\Delta, \quad z = \ln\left(r/R_\odot \right)/\Delta.
\label{eqn:coords}
\end{equation}
In these coordinates, the sizes of each grid cell are $h_x=h_y=\Delta r \sin\theta$ and {$\quad h_z= \Delta r$}, and with our chosen resolution of $\Delta=1^\circ$, the number of grid points is $(361,281,53)$.

To obtain second-order accuracy in the computation of $\bb=\nabla\times\avec$ and $\jj=\nabla\times\bb$, we define the variables on a staggered grid, where $\avec$ and $\jj$ are defined on the cell ribs and $\bb$ is defined on the cell faces.

\subsubsection{Boundary Conditions}
The following boundary conditions are implemented:
\begin{enumerate}
\item{The longitudinal boundaries are periodic for all variables.}
\item{At the latitudinal boundaries, $\theta=10^\circ$ and $\theta=170^\circ$, we set $B_\theta=0$ so that the field is tangential to the boundary.}
\item{The top boundary $r=2.5R_\odot$ is open, and field lines are effectively opened up by the radial outflow velocity.}
\item{On the bottom boundary $r=R_\odot$, corresponding to the photosphere, the evolution of $B_r$ is specified by the surface flux transport model as described in Paper 1. Here the velocity and diffusion take their surface values (see Paper 1), rather than the coronal forms \eqref{velocity} and \eqref{diffusivity}.}
\end{enumerate}

\subsubsection{Initial Condition}
The simulation runs continuously from day of year 106, mid-way through Carrington rotation CR1948, until day 283 at the end of CR1954. The initial condition for $B_r$ on the solar surface for day 106 is obtained from synoptic maps as described in Paper 1. For the initial magnetic field in the coronal volume, we follow the technique of \inlinecite{vanballegooijen2000} to compute the potential field that is periodic in the longitudinal direction, has $B_\theta=0$ on the latitudinal boundaries, has $B_\theta=B_\phi=0$ at the outer boundary $r=2.5R_\odot$, and matches the required distribution of $B_r$ on the solar surface.

The initial potential field is then evolved forward in time between days 106 and 283 using Equation \eqref{induction}, without ever resetting either the photospheric or coronal fields. To maintain accuracy to the observed photospheric field over this period, it is necessary to represent newly-emerging flux in the simulation, and this is described next.

\subsection{Insertion of Magnetic Flux} \label{sec:emerging}

The need for newly emerging flux during the simulation was demonstrated in Paper 1. As we are dealing with many active regions, we cannot represent the flux-emergence process in a detailed, time-dependent fashion. Instead, we must use a simpler technique for inserting magnetic bipoles into the 3D magnetic model. Our approach is to describe the final effect of the newly-emerged active regions on the evolving photospheric and coronal fields.

For each new active region, we use the same mathematical description of a twisted magnetic bipole as \inlinecite{mackay2001} and \inlinecite{mackay4}. This is a 3D field, given in our computational $(x,y,z)$ coordinates by the vector potential
\begin{eqnarray}
A_x &=& \beta B_0 \expt{0.5}z\expt{-2\xi},\\
A_y &=& B_0\expt{0.5}\rho \expt{-\xi},\\
A_z &=& -\beta B_0 \expt{0.5}x\expt{-2\xi},
\end{eqnarray}
where
\begin{equation}
\xi = \frac{\left(x^2+z^2\right)/2 + y^2}{\rho^2}.
\end{equation}
The parameters $B_0$ (peak magnetic field strength) and $\rho$ (half-separation distance between peaks of opposite polarity) are chosen to match observations of each region, as is the tilt angle $\delta$ which simply rotates the field in the $(x,y)$ plane. In the 3D field there is another parameter $\beta$ that alters the twist of the 3D bipole field but does not change the bipole's footprint on the solar surface.

\begin{figure}
  \centerline{\includegraphics[width=0.48\textwidth,clip=]{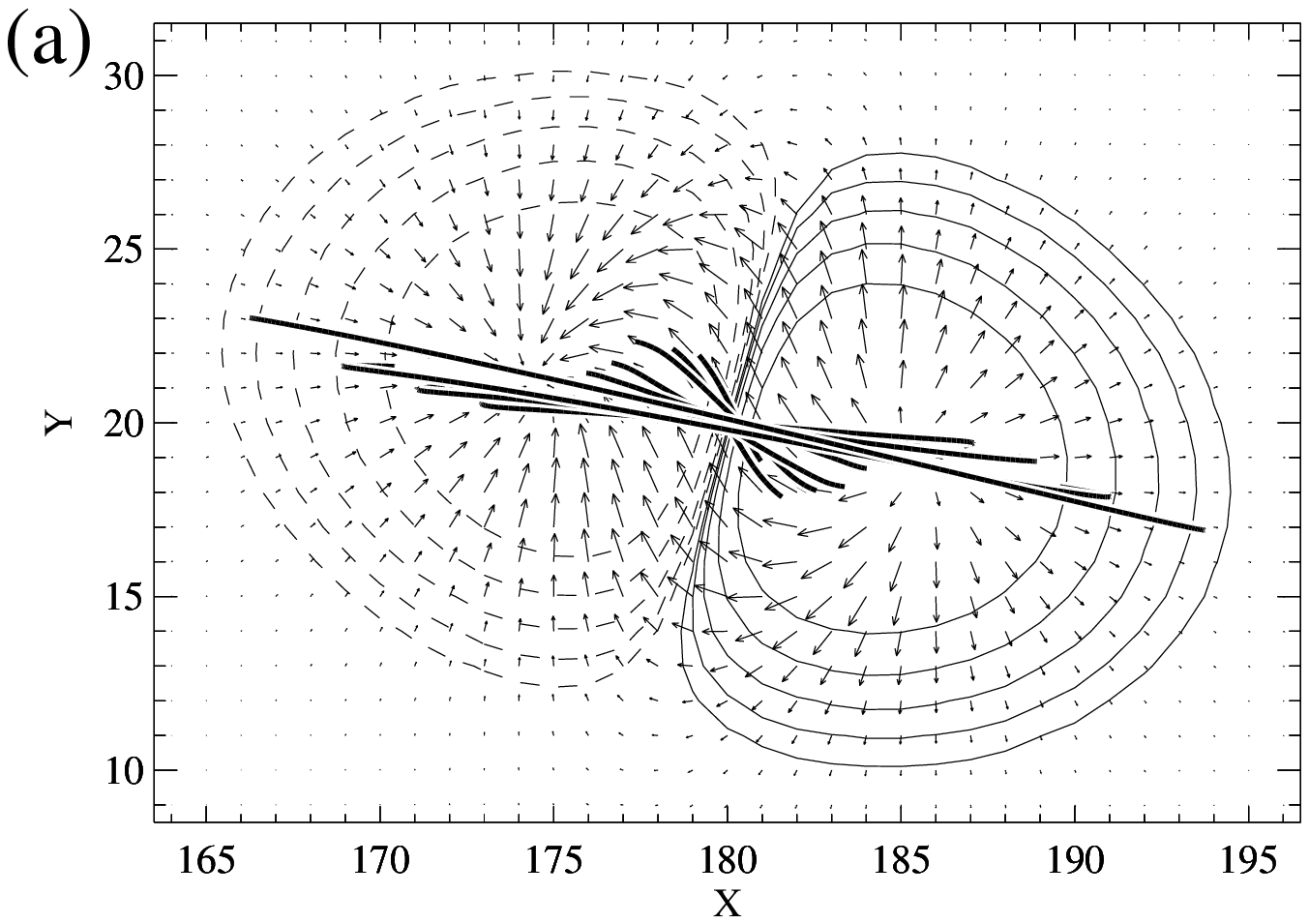}
    \includegraphics[width=0.48\textwidth,clip=]{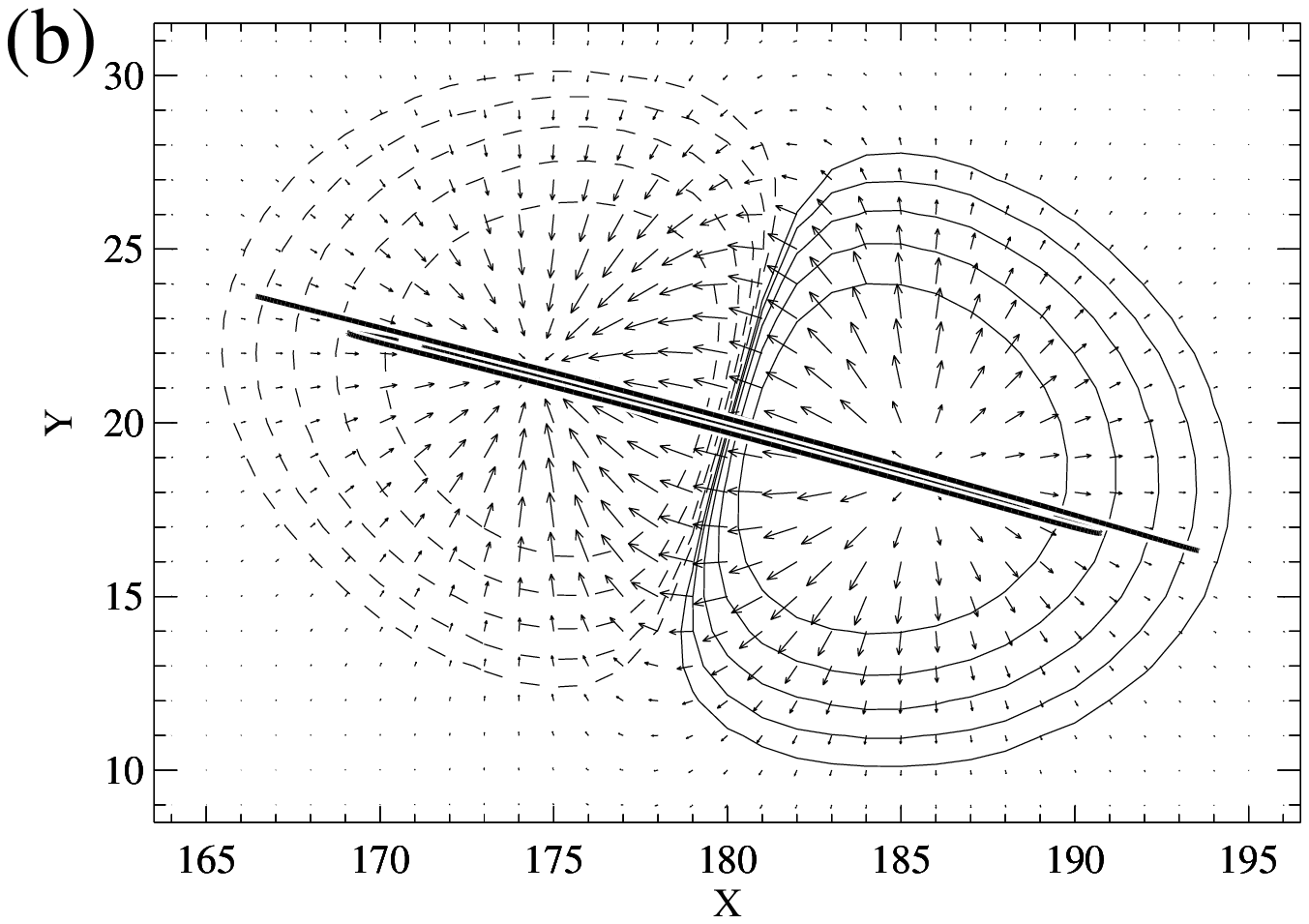}
  }
  \centerline{\includegraphics[width=0.48\textwidth,clip=]{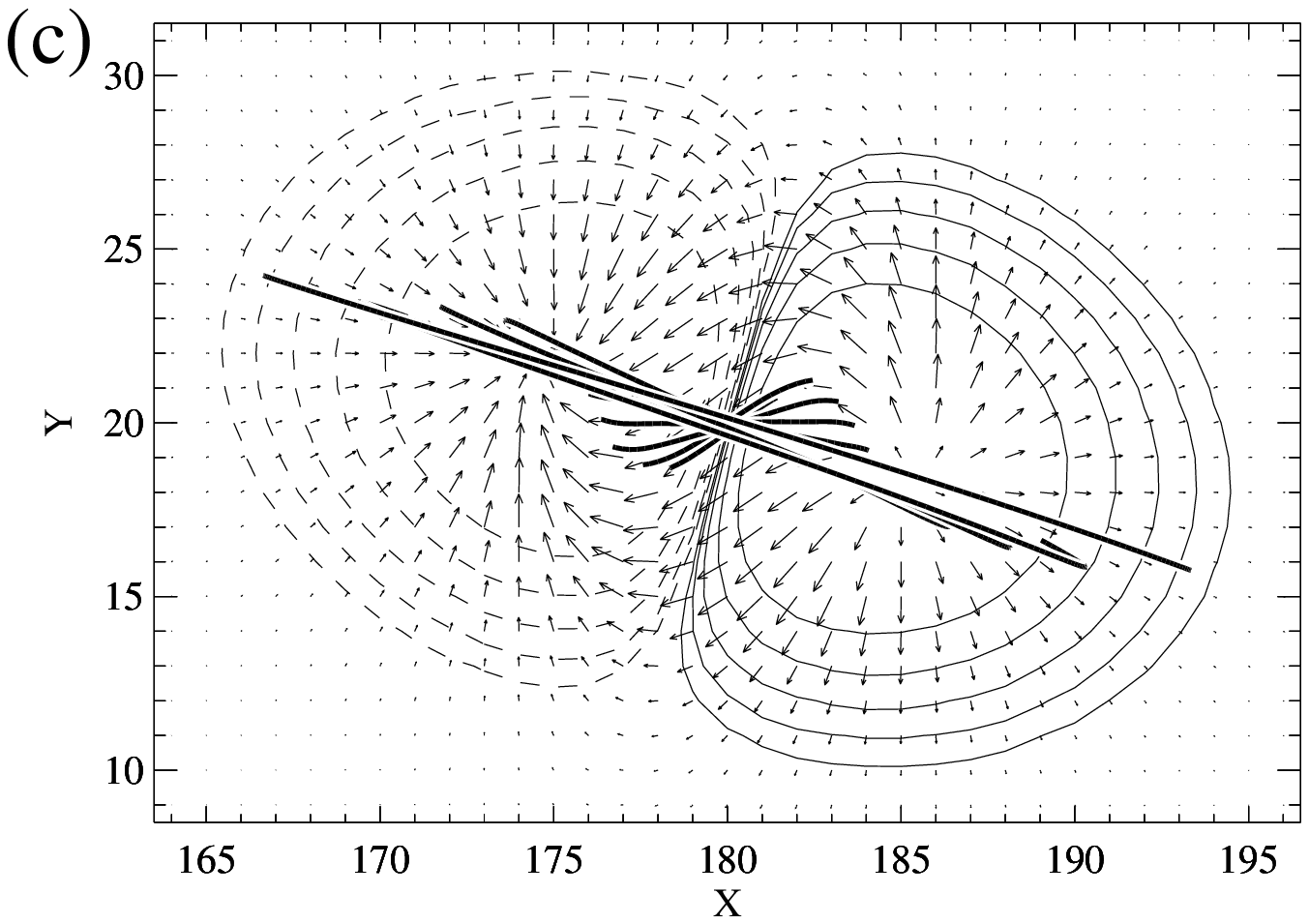}
    \includegraphics[width=0.48\textwidth,clip=]{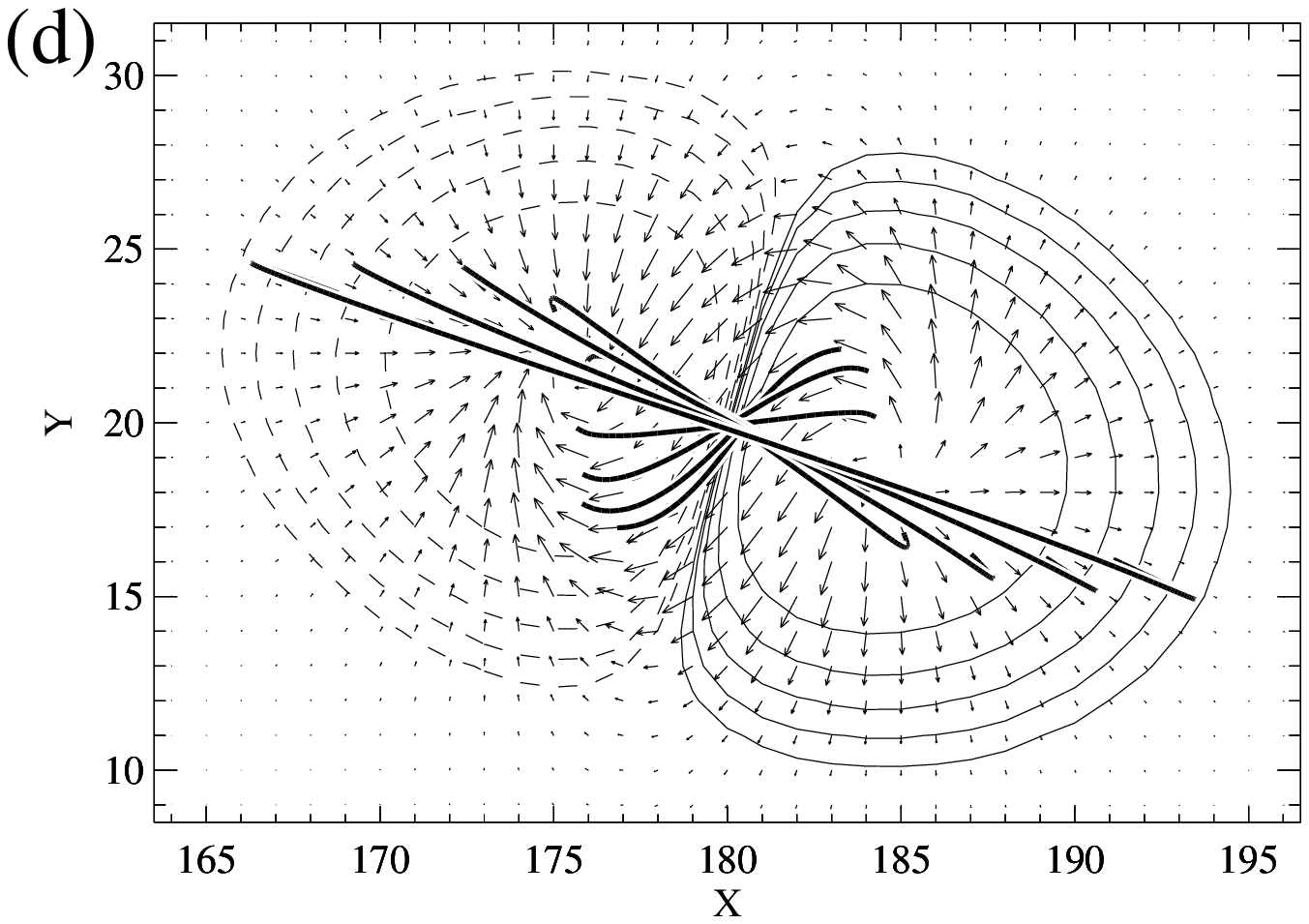}
  }
  \caption{Effect of the twist parameter $\beta$ on a bipolar region: (a) $\beta=-0.2$, (b) $\beta=0.0$, (c) $\beta=0.2$, and (d) $\beta=0.4$. In each case the radial magnetic field on the photosphere is shown by contours (solid for positive flux and dashed for negative flux), and the arrows display the strength and direction of the horizontal components. The thick lines are selected coronal field lines traced from a range of heights above the centre-point of the bipole.}
  \label{fig:bipolebeta}
\end{figure}

The effect of the twist parameter $\beta$ is illustrated by Figure \ref{fig:bipolebeta}, which shows the magnetic field structure for bipoles with four different values of $\beta$. When $\beta=0$ (Figure \ref{fig:bipolebeta}b) the field lines cross from one polarity to the other perpendicular to the polarity inversion line (PIL). A non-zero value of $\beta$, however, alters the angle of field lines with respect to the PIL, as seen in panels (a), (c), and (d) of Figure \ref{fig:bipolebeta}. This angle varies with height in the corona, and is greatest for the lowest field lines, nearest to the bipole centre. The resulting twist corresponds to a non-zero magnetic helicity; for our domain the appropriate definition would be the relative helicity \cite{berger1984,finn1985}. However, in this paper we are interested primarily in the skew direction at the particular locations of observed filaments, and we do not explicitly calculate any helicity integrals. It can be shown that the sign of $\beta$ matches the overall sign of relative helicity for each bipole, and this sign also corresponds to the skew direction of field lines above the central PIL. For negative $\beta$ (Figure \ref{fig:bipolebeta}a) the skew is dextral, while for positive $\beta$ (Figures \ref{fig:bipolebeta}c and \ref{fig:bipolebeta}d) the skew is sinistral. We will discuss the choice of $\beta$ in different simulation runs in Section \ref{sec:parameters}.

For the 6-month period considered in this paper, 119 bipolar regions were identified in Paper 1 from synoptic magnetogram observations. For each region we recorded the day of observation, bipole-centre coordinates, tilt angle, half-separation distance between peaks, and magnetic flux. Magnetic bipoles matching the properties of the observed regions are inserted into the simulation at the appropriate days. In fact, as explained in Paper 1, each bipole is inserted into the simulation 7 days before its date of observation at central-meridian passage, with its properties ``evolved'' back in time for 7 days. This arbitrary choice of 7 days and its consequences are discussed in Section \ref{sec:parameters}.

\begin{figure}
  \centerline{\includegraphics[width=0.4\textwidth,clip=]{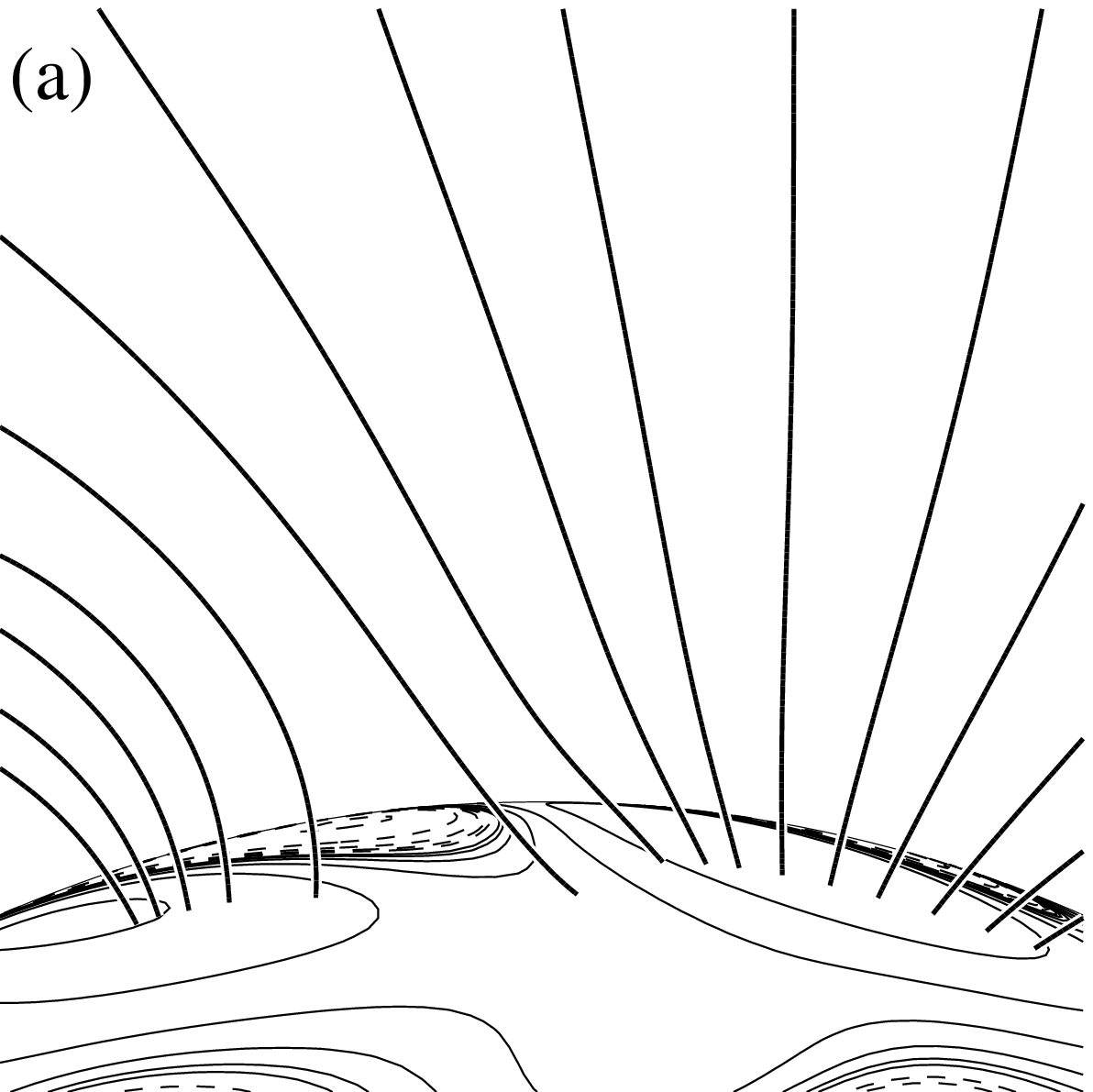}
    \hspace*{0.05\textwidth}
    \includegraphics[width=0.4\textwidth,clip=]{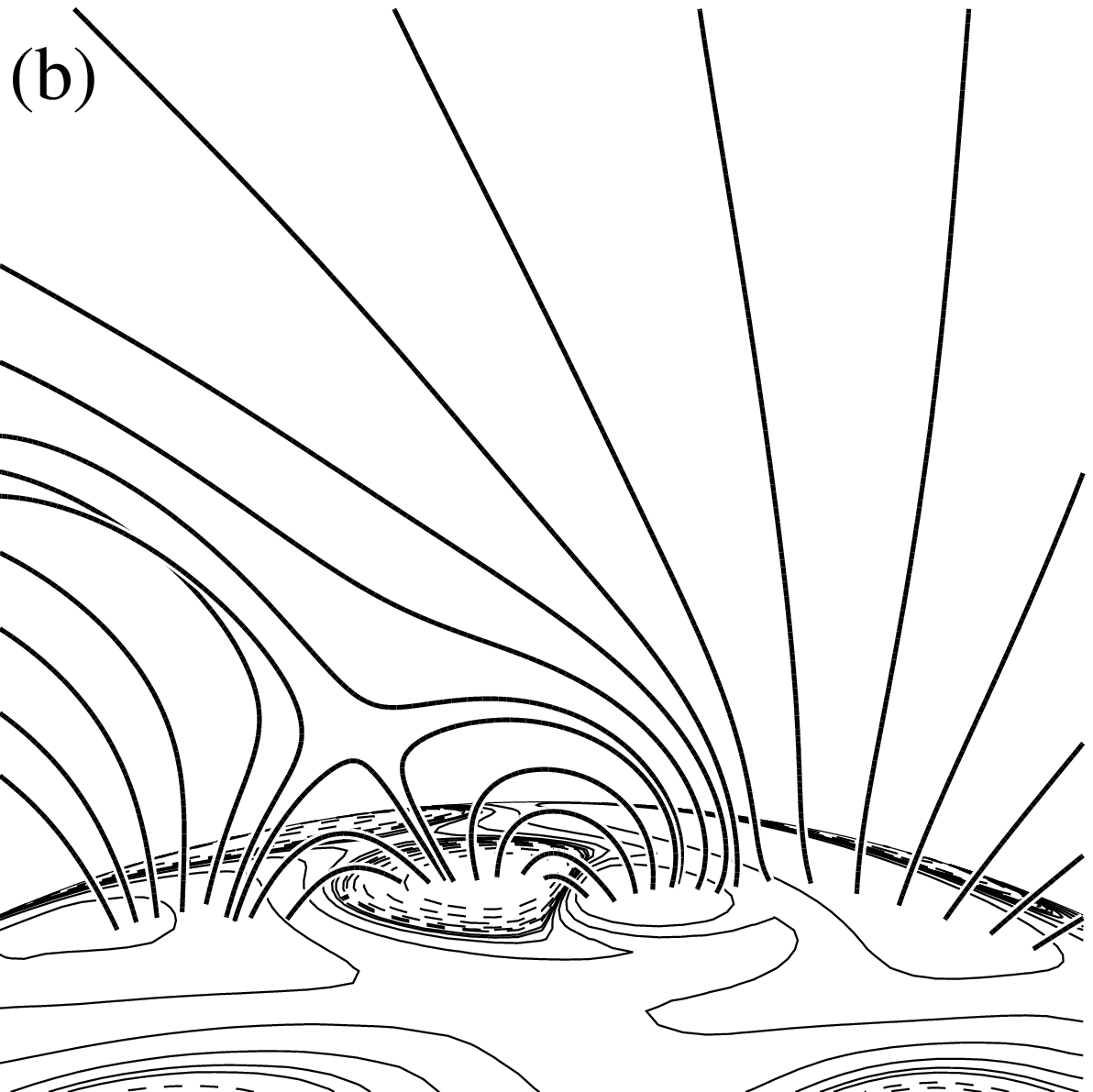}
  }
  \caption{Example insertion of a bipolar region, seen ``before and after'' on (a) day 250 and (b) day 251. Contours on the surface show radial magnetic field strength, with solid lines for positive flux and dashed lines for negative flux. Thick lines represent selected coronal magnetic field lines.}
  \label{fig:insert3d}
\end{figure}

The bipole insertion is instantaneous, but any pre-existing magnetic field in the insertion region is ``swept'' away before the insertion. This models the expected distortion of pre-existing coronal field by a newly-emerging flux tube, as seen in simulations of flux emergence \cite{yokoyama1996,krall1998}. It also prevents the formation of disconnected magnetic flux in the corona. The insertion process is summarised as follows:
\begin{enumerate}
\item{Simulation time is frozen during the insertion process, and the ordinary flows and diffusion are turned off in both the photosphere and corona.}
\item{An outward velocity of the form $v_s=\exp\left(-As\right)$ is applied in the photospheric and coronal region surrounding the bipole centre, where the coordinate $s$ represents distance from the bipole centre. The constant $A$ is chosen so as to give zero flow at the edge of the insertion region. This sweeping advects the field out of the insertion region, and sweeping stops once the field near the bipole centre reaches a suitably low level. We have found that a threshold of $0.05B_0$ works well, where $B_0$ is the maximum magnetic field of the new bipole. To ensure numerical stability, a localised photospheric diffusion is applied in the insertion region during this process.}
\item{The new bipole is inserted by adding the corresponding vector potential to the (weak) existing field within the insertion region.}
\item{The coronal field in the bipole region is then allowed to relax toward equilibrium for 50 time steps. The coronal diffusion is turned on again during this process so as to allow the bipole to reconnect with its surroundings.}
\end{enumerate}
An example of the coronal field resulting from such a bipole insertion is shown in Figure \ref{fig:insert3d}, for the same bipolar region as in Figure 8 of Paper 1. Notice how the pre-existing field on day 250 (Figure \ref{fig:insert3d}a) has largely been swept outside the new bipolar region on day 251 (Figure \ref{fig:insert3d}b). The newly-emerged bipole has reconnected to create a more complex topology, with two new closed flux domains connected to the new bipole. The field at a larger distance is little altered however.

\subsection{Simulation Parameters} \label{sec:parameters}
Our method for the insertion of newly emerging active regions is necessarily artificial, due to the lack of ideal observational data at the present time. There are two key parameters that contain these uncertainties in our model, namely the date of insertion and the twist parameter $\beta$ of each bipole.

As we can only observe one side of the Sun, the exact date of emergence of each active region cannot be determined from the available observations. It is therefore necessary to choose an arbitrary date of insertion of each bipole in the model, which we take to be some fixed time before the region reached central-meridian passage as recorded in the synoptic magnetograms. As explained in Paper 1 we have chosen this time to be 7 days for all regions. With our ``backward-evolution'' of bipole parameters (Appendix 2 of Paper 1), we have found that the number of days chosen does not particularly affect the surface field. It does slightly change the coronal field, as inserting bipoles earlier or later will increase or decrease the skew angle of the coronal field produced on a particular day. However, the overall chirality type (\emph{i.e.} dextral, sinistral, or weak skew), which is the focus of this study, will not be affected.

The twist parameter $\beta$ for each new bipole is a key component of the model that introduces helicity into the coronal field. In principle the helicity of each active region should come from detailed vector magnetograph data in 3D space, which is unfortunately not available. As the available synoptic magnetograms give only the normal magnetic field component $B_r$ we are unable to determine the twist of each region. Due to this limitation, we have assumed for simplicity that all regions in each hemisphere have the same $\beta$ value. Note that this does not mean that the magnetic helicity is constant in space in our model. After a bipole has been inserted into the 3D model and the coronal field has relaxed to a force-free state, it is possible to compute the ``current helicity'' $\alpha(\vec{x})$, which is a measure of the \emph{local} twist of field lines, and is defined by $\nabla\times\bb = \alpha\bb$. However, this function $\alpha(\vec{x})$ is not directly related to our parameter $\beta$. The current helicity $\alpha$ is concentrated mostly towards the centre of the bipole, and is \emph{not} uniformly distributed in space. Choosing the same $\beta$ for all bipoles in a given hemisphere does not result in $\alpha$ being constant in space.

The effects of different $\beta$-values on the coronal field evolution have been considered in previous simulations. For example, \inlinecite{mackay4} considered the interactions of two bipoles with varying degrees of twist, showing how the twist evolved through the corona over a 60-day period. Varying the value of $\beta$ affected the length of time for flux ropes to develop in the corona above each bipole, but not the overall chirality type, provided that the sign of $\beta$ was not changed. That paper was designed as a simple simulation representing characteristic interactions which will be occuring in the present simulation. Here we have run four simulations with different values of $\beta$ for the emerging bipolar regions, so as to consider the effect of emerging active region helicity on filament chirality.

The four simulations are summarised in Table \ref{tab:twists}. They include runs where the sign of $\beta$ follows the observed hemispheric sign for helicity (runs 3 and 4), and a run where the sign of $\beta$ in each hemisphere is opposite to that observed (run 1). There is also a run where the inserted bipoles are untwisted (run 2). Based on the results of \inlinecite{mackay3}, we expect this active-region helicity to have an important effect on the magnetic skew at filament locations, with the correct hemispheric pattern only produced for the correct hemispheric sign of bipole twist $\beta$. In the next section we test this hypothesis by comparing our four simulation runs with the observations of real filaments.

\begin{table}[ht]
\caption{Summary of bipole twist used in different simulation runs.}
\label{tab:twists}
\begin{tabular}{clll} \hline
Simulation run & N hemisphere & S hemisphere & Hemispheric sign \\
\hline
(1) & $\beta=0.2$ & $\beta=-0.2$ & opposite to observed\\
(2) & $\beta=0$ & $\beta=0$ & none \\
(3) & $\beta=-0.2$ & $\beta=0.2$ & as observed\\
(4) & $\beta=-0.4$ & $\beta=0.4$ & as observed \\
\hline
\end{tabular}
\end{table}

\section{Skew Results} \label{sec:skew}

To compare simulations to observations, we use the filament data set described in Paper 1, comprising daily H$\alpha$ observations from BBSO for 255 filaments. Of these, the chirality could be clearly determined for 123 filaments, as outlined in Paper 1. In this section, we compare the chirality of 109 individual filaments in the sample with the skew of magnetic field lines at corresponding PIL locations in the simulation. This is a valid comparison because observations show a one-to-one correspondance between the axial magnetic field within filaments and in the surrounding corona \cite{martin1998}. As an example, Figure \ref{fig:filament544} illustrates a single filament from the sample. In the simulated magnetic field (Figures \ref{fig:filament544}a and \ref{fig:filament544}c) there is a clear flux rope low in the corona, with an overlying sheared coronal arcade. The field is seen to be dextrally skewed, matching the observed filament, which has dextral barbs visible in H$\alpha$ (Figure \ref{fig:filament544}b) and must therefore have a dextral axial field \cite{martin1998}. The simulation thus reproduces the correct chirality for this filament.

Of the 255 observed filaments in the sample, the simulation was found to reproduce the local photospheric PIL shape accurately in 207 cases. The remaining 48 observed filaments have no corresponding PIL in the simulation so have been omitted from the results in this paper. In Section \ref{sec:pattern} we use the 207 simulated filament locations to look at the overall proportions with dextral and sinistral skew in the simulation. The main result of this study is then presented in Section \ref{sec:indiv}, where we compare the skew at individual filament locations with the chirality of the filament observed there. This is done for the 109 locations where we both determined the chirality of the observed filament clearly (Paper 1), and simulated the PIL shape accurately.

\subsection{Measurement of Skew}

The skew of the simulated magnetic field at each of the 207 simulated filament locations was determined using the following technique.

We first define the skew angle $\gamma$ to be the angle between a horizontal line normal to the PIL, and the horizontal magnetic field at a given height above the photosphere \cite{mackay2001}. This skew angle will vary at different heights, and also with distance along the PIL. For each filament location we measured the skew at three heights, $z=1,2$, and $3$, corresponding to heights of $14\megam$, $28\megam$, and $35\megam$ above the photosphere. We computed the fractional lengths of dextral skew ($\gamma > 30^\circ$), sinistral skew ($\gamma < -30^\circ$), and weak skew ($|\gamma| < 30^\circ$), along the part of the PIL where the filament was observed.

To determine the overall skew of the simulated filament location, the fractional lengths of each skew type  were first averaged over the three heights. Then the filament location was classified as dextral or sinistral according to which average length was greater. If the fractional length of weak skew was greater than 0.8 at all heights then the filament location was classified as having weak skew.

\subsection{Simulated Hemispheric Pattern} \label{sec:pattern}

Before comparing individual filaments in the next section, we consider the overall number of dextral and sinistral locations in each hemisphere of our simulation.

\begin{figure}
  \centerline{\includegraphics[width=0.9\textwidth,clip=]{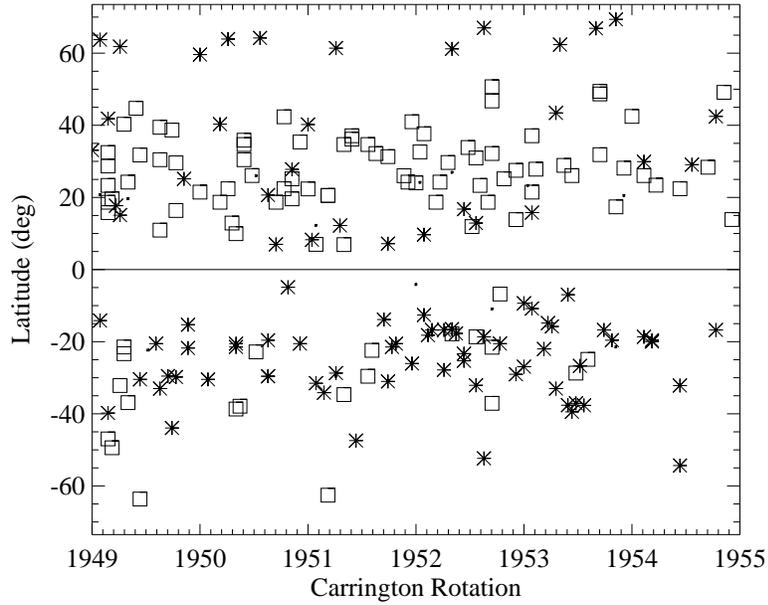}
  }
  \caption{Skew of the 207 simulated filament locations for simulation run 3, on a time-latitude plot. Squares show locations with dextral skew, asterisks show locations with sinistral skew, and small dots show locations with weak skew.}
  \label{fig:latrotlarge}
\end{figure}

The classification of each of the 207 filament locations is shown in Figure \ref{fig:latrotlarge}. This is for simulation run 3 in Table \ref{tab:twists}, {\it i.e.} with twist $|\beta|= 0.2$ and the correct hemispheric sign of helicity. In summary there are 100 dextral locations (squares on the figure), 94 sinistral locations (asterisks), and 13 locations with weak skew (small dots). This plot should be compared with the observed filament chiralities in Figure 4 of Paper 1.

If we discount the 13 filament locations with weak skew, then 71.2\% of filament locations in the Northern hemisphere are dextral, while 74.7\% in the Southern hemisphere are sinistral. In simulation run 4, which still has the correct hemispheric sign of twist, but an increased magnitude of $|\beta|=0.4$, these percentages increase to 72.3\% in the Northern hemisphere and 80.2\% in the Southern hemisphere. They agree roughly with our observed filament sample, where the respective percentages are 88.7\% and 73.1\%, and also with previous observational results \cite{martin1994,pevtsov2003}. One clear feature in Figure \ref{fig:latrotlarge} is that all of the high-latitude filaments (above $60^\circ$) in our simulation have the opposite skew to the hemispheric pattern. This will be explained in Section \ref{sec:discussion}.

\subsection{Comparison of Individual Filament Chirality} \label{sec:indiv}

We now move to the direct comparison of individual observed filament chiralities with the simulated skew. All 109 observed filaments used for the comparison are classified as either dextral or sinistral. Table \ref{tab:correct} shows the percentage with correctly reproduced chirality, for each simulation run. There is a clear improvement in the simulations with the correct sign of bipole twist in each hemisphere (namely runs 3 and 4). Moreover, the simulation with stronger twist, run 4, performs better than run 3, correctly classifying 86.2\% of filaments. The percentages in Table \ref{tab:correct} are overall results for the whole sample, but we have been able to explain many of the disagreements in this classification by more careful analysis of the individual filaments. 

\begin{table}[ht]
\caption{Percentage of filaments with correctly reproduced chirality, in each simulation run.}
\begin{tabular}{lll} \hline
Simulation run & \hspace{1cm} &  Percentage correct  \\
\hline
(1) \quad wrong sign of $\beta$ && 61.5\% \\
(2) \quad $\beta=0$ (no bipole twist) && 70.6\% \\
(3) \quad $|\beta|=0.2$, correct sign && 83.5\% \\
(4) \quad $|\beta|=0.4$, correct sign&& 86.2\% \\
\hline
\end{tabular}
\label{tab:correct}
\end{table}

\begin{figure}[hp]
  \centerline{\includegraphics[width=0.82\textwidth,clip=]{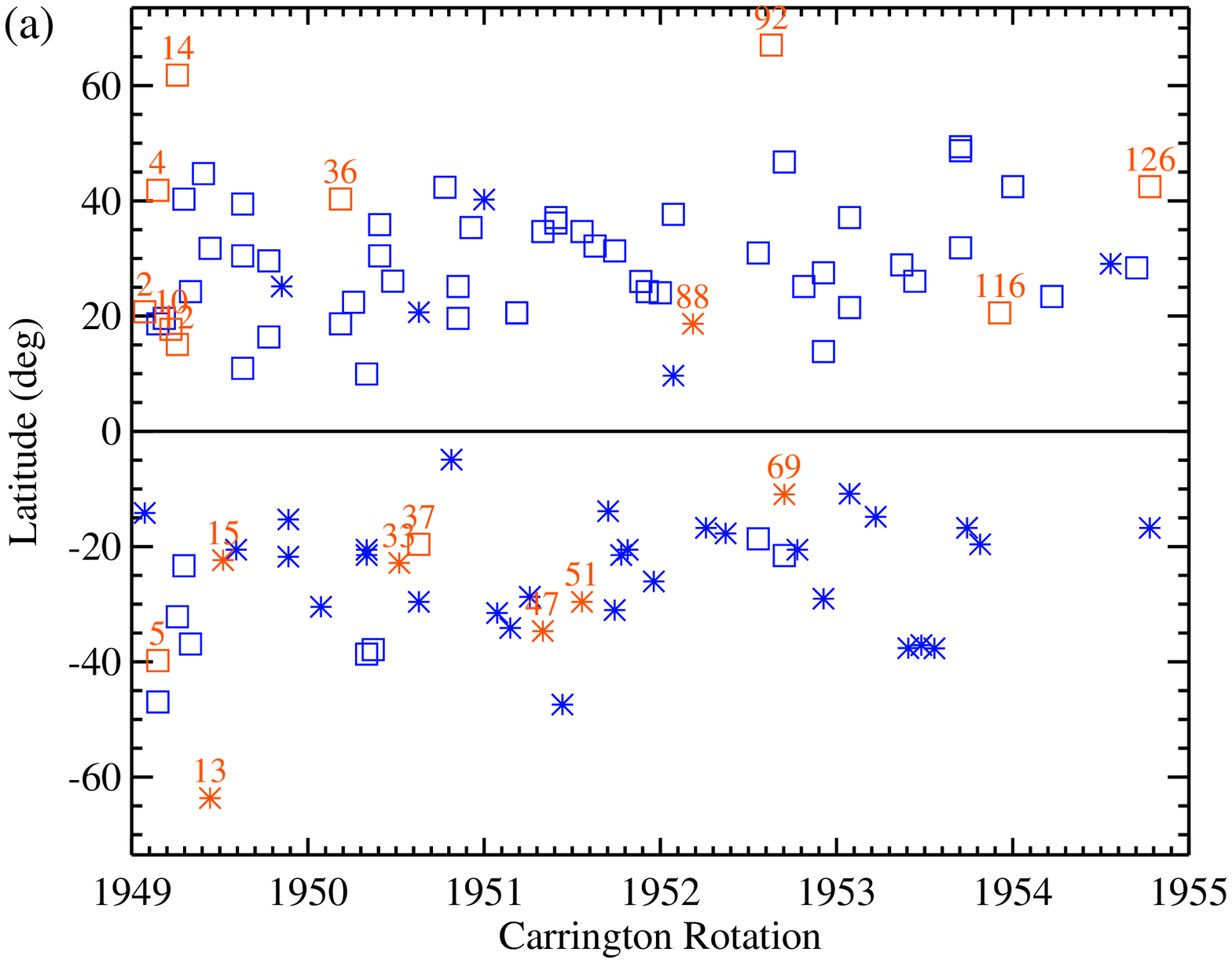}
  }
  \centerline{\includegraphics[width=0.82\textwidth,clip=]{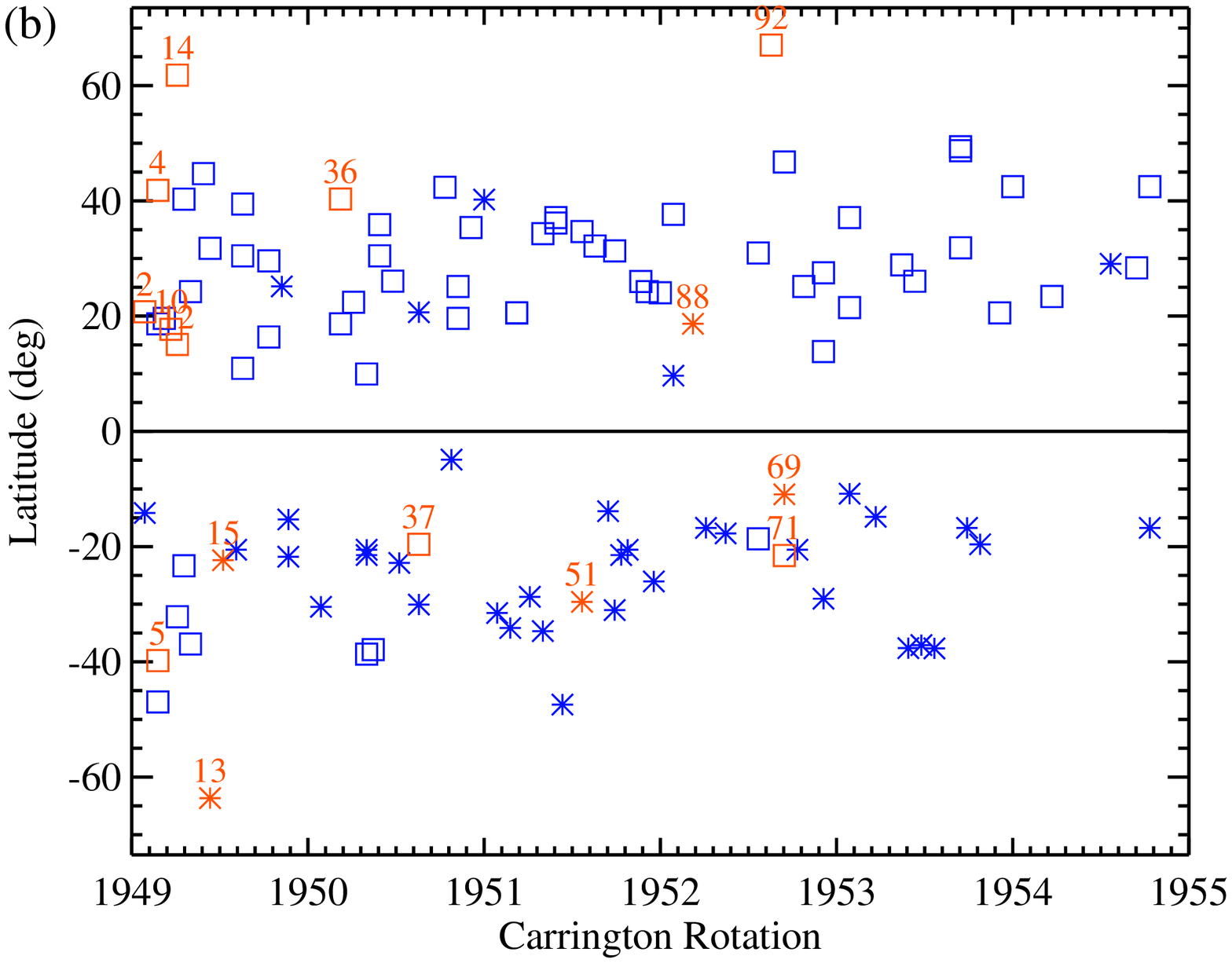}
  }
  \caption{Results of the chirality comparison for the simulation runs with correct hemispheric pattern of helicity, (a) run 3 with $|\beta|=0.2$, and (b) run 4 with $|\beta|=0.4$. Shapes denote chirality of observed filaments, with squares for dextral and asterisks for sinistral. Colours show whether simulated skew matches observed chirality, blue meaning correct and red incorrect (with filament ID numbers).}
  \label{fig:correctskew}
\end{figure}

\begin{figure}[hp]
  \centerline{\includegraphics[width=0.82\textwidth,clip=]{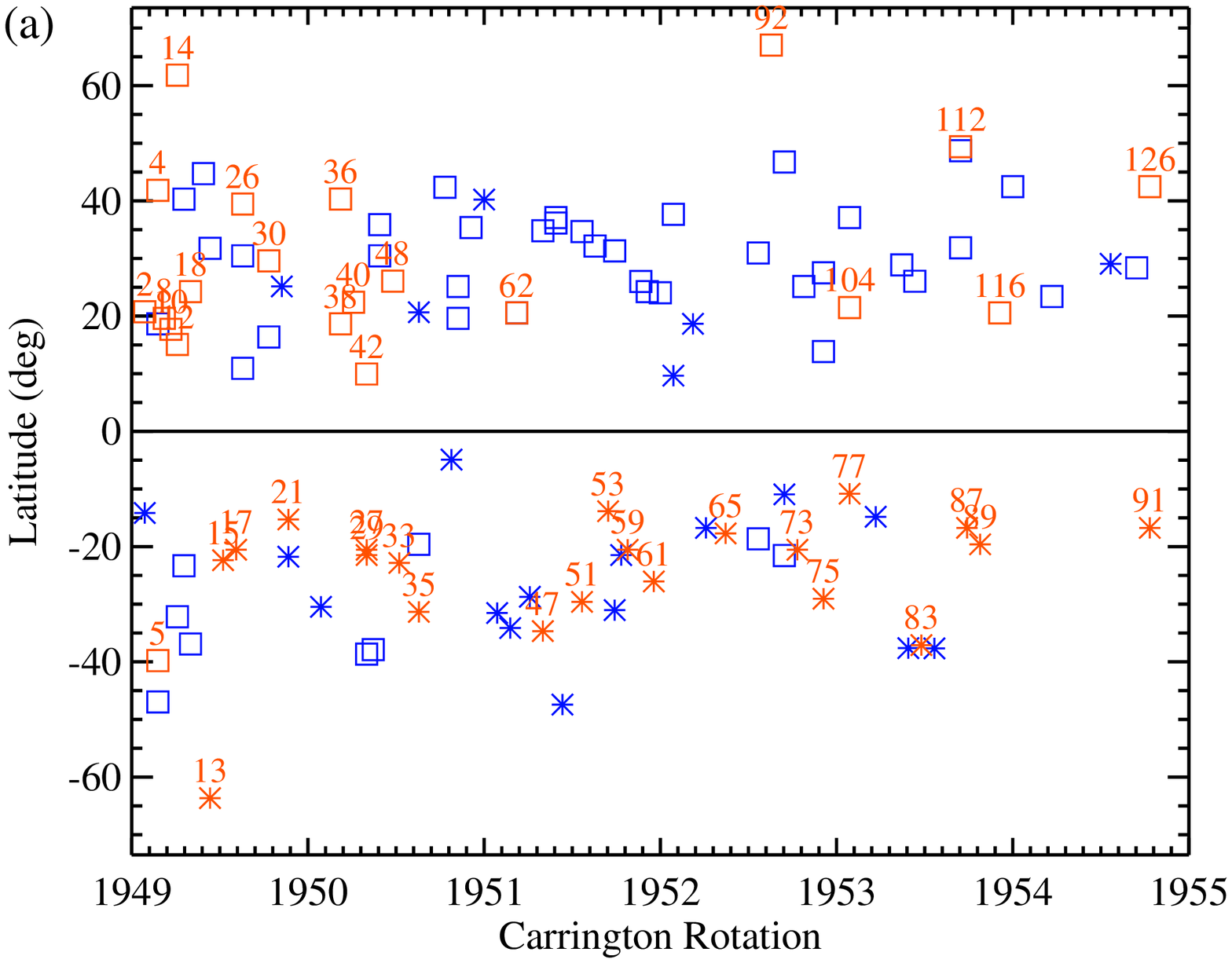}
  }
  \centerline{\includegraphics[width=0.82\textwidth,clip=]{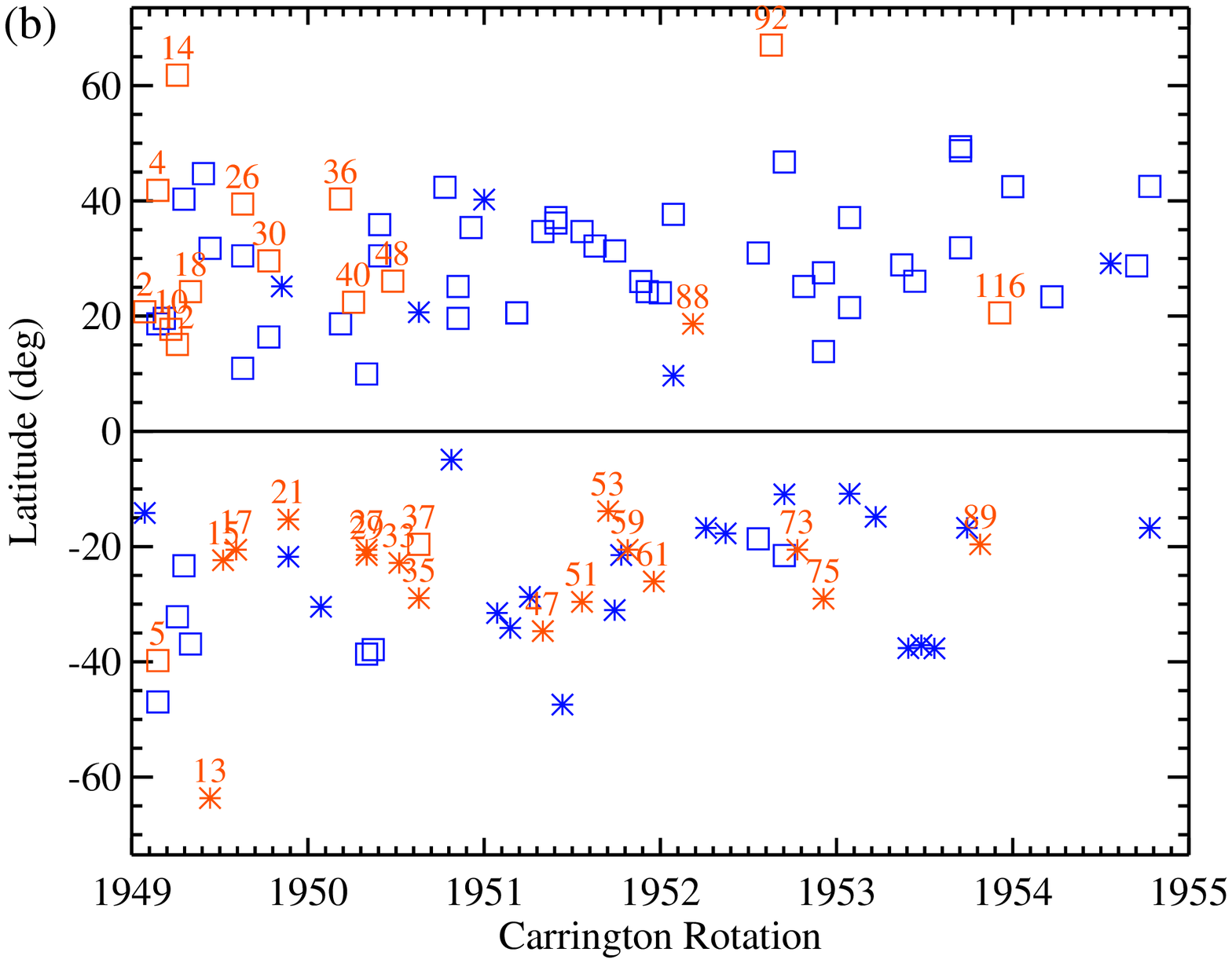}
  }
  \caption{Results of the chirality comparison for (a) run 1 where the sign of twist is opposite to that observed, and (b) run 2 where the bipoles have no twist ($\beta=0$). Shapes denote chirality of observed filaments, with squares for dextral and asterisks for sinistral. Colours show whether simulated skew matches observed chirality, blue meaning correct and red incorrect (with filament ID numbers).}
  \label{fig:correctskew2}
\end{figure}

From Figures \ref{fig:correctskew} and \ref{fig:correctskew2} we can see the distribution of wrongly classified filaments over both latitude and time. The figures show individual filament results on time-latitude plots for the four simulations. For each filament, the observed chirality is shown by the shape, with squares for dextral and asterisks for sinistral. The colour shows whether the simulated skew at that location matches the observed filament chirality. Blue means the simulation is correct, and red means it is wrong. In addition, the filaments with wrong skew have an identification number printed on the plot next to the red symbol. These numbers are increasing in time, even in the Northern hemisphere, and odd in the Southern hemisphere.

In simulation runs 3 and 4, with the correct hemispheric pattern of bipole twist (Figure \ref{fig:correctskew}), we see a general distribution of wrongly classified filaments across all latitudes, but an improvement in the classification over time. In Section \ref{sec:discussion} the individual circumstances of the wrongly classified filaments are examined, and in most cases the disagreement can be explained. Figures \ref{fig:correctskew2}(a) and (b), corresponding to runs with either the wrong sign of bipole twist (run 1) or no bipole twist (run 2), show poorer results throughout the simulation, as expected.

An important result concerns those filaments which are exceptions to the hemispheric chirality pattern. In the simulations with correct sign of bipole twist, we see from Figure \ref{fig:correctskew} that the minority type of skew is correctly produced for the majority of such filaments. The model is able to reproduce not just the general hemispheric pattern, but also the variations in individual circumstances which lead to exceptions. This is strongly suggestive that the model contains the correct physics, and supports the view that filament chirality results from the properties of active regions, their development, and their interaction as they decay and are transported poleward. 

\section{Discussion} \label{sec:discussion}

The results of the comparison in this paper are very encouraging; the model is able to reproduce the correct skew for the majority of filaments, even in the rather complex global magnetic field. In this section we look at why the model fails to produce the correct chirality in all cases, and find that there are certain filaments which we could not expect to match correctly in the present simulation. If these filaments are removed from the comparison then the results are considerably improved, up to 96.9\% correct for simulation run 4.

Figures \ref{fig:correctskew}(a) and (b) show that a large number of disagreements occur near the beginning of the simulation, in the first two weeks of Carrington rotation CR1949. This may readily be explained by the initial condition used in the coronal model, a potential field. This initial field has no magnetic shear, and so the skew is initially weak over all PILs. Furthermore, over the 8 day ``ramp-up'' time before the first bipole insertion, the action of differential rotation tends to produce the wrong direction of skew in each hemisphere \cite{leroy1978}. This explains why the wrong skew is developed for 7 of the lower-latitude filaments early in the simulation (numbers 2, 4, 5, 10, 12, 15, and 36). They all form above PILs bounded by magnetic flux from the initial potential field. There are also filaments early in the simulation with the correct skew, but these are associated with newly emerged bipolar regions. Once a significant amount of new flux has emerged and the field has evolved away from the initial configuration, a much better agreement is found. This indicates that the long-term development of the coronal field is important and should be included in simulations, rather than taking separate extrapolations at different times.

The initial condition also accounts for the three higher-latitude filaments included in the comparison, numbers 13, 14, and 92 on Figures \ref{fig:correctskew} and \ref{fig:correctskew2}. These are all wrongly classified as having the minority chirality for their hemisphere. Although filament 92 is observed almost 4 months into the simulation, the surrounding field is still a relic of the initial potential field. This is because the relatively slow poleward meridional flow (peak speed $16\mpsec$) takes about two years to transport magnetic flux from the active belts to the polar regions. In Figure \ref{fig:latrotlarge}, we see that the wrong hemispheric sign of skew is found at the northern polar crown until at least the end of CR1953 in our simulation. We could not expect to reproduce the correct chirality for such high-latitude filaments unless we run the simulation for a much longer period.

Table \ref{tab:wrong} summarises the numbers of wrongly classified filaments for each simulation run, broken down into five different classes. The first two classes, ``polar crown'' and ``early'', correspond to the filaments described above. These classes have the same number of filaments for each run, as they are caused by the initial conditions and are not affected by the twist of newly emerging regions. The other three classes show different numbers for different runs, and account for the improved totals in simulations with the correct sign of bipole twist.

\begin{table}[ht]
\caption{Classification of filaments with wrongly simulated chirality, for each simulation run.}
\begin{tabular}{cllllll} \hline
Simulation run & Polar crown & Early & Internal PIL & Weak skew & Other & Total \\
\hline
(1) &  3 & 7 & 17 &  4 & 11 & 42\\
(2) &  3 & 7 &  4 & 10 &  8 & 32\\
(3) &  3 & 7 &  2 &  2 &  4 & 18\\
(4) &  3 & 7 &  2 &  2 &  1 & 15\\
\hline
\end{tabular}
\label{tab:wrong}
\end{table}

The class ``internal PIL'' contains those filaments which form above PILs lying within a single bipolar region. They have traditionally been known as Type A \cite{mackay1}. The simulated skew at these locations is directly affected by the twist $\beta$ of the emerged bipolar region, which in each of our simulations has the same value for all bipoles in a single hemisphere. As illustrated by Figure \ref{fig:bipolebeta}, negative $\beta$ produces dextral skew and positive $\beta$ produces sinistral skew. This direct effect of the bipole twist explains the wrong skew of filaments 37 and 88 in the simulations with the correct hemispheric signs of $\beta$ (runs 3 and 4, shown in Figure \ref{fig:correctskew}). These are both Type A filaments which do not follow the chirality pattern, implying that the active regions concerned probably had the opposite sign of helicity to the majority in their hemisphere. When we change the sign of bipole twist in the simulation (run 1), we find 17 Type A filaments with the wrong skew, but filaments 37 and 88 now have the correct skew.

In each simulation run there were a small number of filament locations without significant dextral or sinistral skew. These are labelled ``weak skew'' in Table \ref{tab:wrong}. Unsurprisingly they are most numerous in the simulation with zero bipole twist.

The final class of wrongly simulated filaments, labelled ``other'' in Table \ref{tab:wrong}, consists only of filaments forming between flux from different bipolar regions, as Type A filaments have already been accounted for. The number of filaments in this category shows a marked decrease from one simulation run to the next, moving down the table. This illustrates the importance of the emerging bipole twist, even for filaments not internal to a single bipole, and supports the earlier findings of \inlinecite{mackay3}. For the simulations with the correct sign of twist (runs 3 and 4), there is an improvement when the strength of the twist is increased from $|\beta|=0.2$ to $|\beta|=0.4$. In the earlier simulations with a pair of bipoles, \inlinecite{mackay4} demonstrated how flux ropes above the PIL between the two bipoles formed more quickly when the twist strength was increased, although they did form eventually even if the bipoles were untwisted. Our present simulations indicate that the correct sign of bipole twist assists the development of magnetic skew on a fast enough timescale to explain the chirality of observed filaments.

\subsection{Revised Statistics}

The overall performance of the model, in the simulations with correct sign of bipole twist, is very good. Given the limitations of the present simulation, as described above, it would seem reasonable to omit filaments in the first three columns of Table \ref{tab:wrong} from the comparison. In that case the percentage of filament locations with correct skew is 93.8\% for run 3 (6 mistakes), and increases to 96.9\% for run 4 (3 mistakes) when the strength of twist is increased.

\section{Conclusion} \label{sec:conclusion}

We have developed new simulations of the large-scale magnetic field of the solar corona, over a period of many months. Based on the flux transport and magnetofrictional model of \inlinecite{vanballegooijen2000}, the coronal field evolves in response to flux emergence and the changing photospheric boundary conditions. Using a reduced form of the MHD equations, the corona evolves in essence through a continuous sequence of force-free equilibria. In Paper 1 we described our technique for incorporating newly-emerging active regions based on observations, and showed how the large-scale photospheric field remains accurate over a 6-month period. In this paper we have demonstrated how our 3D model is able to reproduce an important observed property of filament magnetic fields, namely the hemispheric chirality pattern. By direct comparison with H$\alpha$ observations of real filaments over a 6-month period in 1999, we have shown that the emergence, surface transport, and reconnection of large-scale twisted active-region magnetic fields are sufficient to produce the chirality pattern.

Our simulations represent a significant improvement over coronal field models where independent extrapolations are taken at different times, particularly those which assume a potential field in the corona. In particular, the newly emerged fields in our simulation are assumed to be twisted, injecting magnetic helicity into the solar atmosphere. The high electric conductivity of the coronal plasma causes magnetic helicity to be conserved \cite{priest2000}, and gives the coronal field a certain degree of memory. This memory is retained because the simulation is a continuous evolution in which the magnetic field is never reduced to its lowest energy state. This allows helicity to be transported from low to high latitudes as the coronal field evolves in response to photospheric motions. We believe such helicity transport to play a key role in filament chirality; if high latitude quiescent filaments were formed \emph{in situ} by differential rotation of a preexisting coronal field then their chirality would be opposite to the observed pattern.

For each filament in our sample of 109 with definite chirality (as inferred from the H$\alpha$ observations--see Paper 1 for details), we have compared the local skew direction of the simulated coronal magnetic field with the observed filament chirality. We find the following main results:
\begin{enumerate}
\item{The simulation produces the correct skew for up to 96.9\% of filaments, including exceptions to the hemispheric pattern. This is for simulations where the sign of newly-emerging bipole twist matches the observed hemispheric sign of active-region helicity \cite{pevtsov1995}.}
\item{If the emerging regions have zero twist, or their sign of twist is opposite to the observed pattern, then the percentage of filaments with correct skew is significantly lower.}
\item{The simulation produces a higher proportion of filaments with correct chirality at later times.}
\end{enumerate}

The results suggest that the four ingredients of differential rotation, meridional circulation, surface diffusion, and emerging twisted active regions are sufficient to produce the observed filament chirality pattern. The results presented in this paper give a comparison of the chirality at a large number of filament locations, and we have not discussed the circumstances of individual locations in any detail. Previous papers using the simpler configuration of two interacting bipoles (\opencite{mackay4}, \citeyear{mackay6}) have considered in detail the basic physical interactions which occur in the model. Very similar features are found at multiple locations in the global simulations presented in this paper.

In addition to forming the necessary sheared magnetic fields, the non-potential simulations have allowed us to vary the twist of the emerging regions, and our results support the conclusions of \inlinecite{mackay3} that this active-region helicity has an important influence on filament chirality. It is not however the only factor; even the simulation run with bipole twist opposite to the observed pattern produced the correct chirality for over 60\% of filaments.

The third result, namely that the performance of the simulation improves with time, is a key conclusion of this study. A large number of wrongly-classified filaments were early in the simulation, before magnetic helicity could build up from the shearing and evolution away from the initial potential field, and from the emergence of new bipoles. This highlights the importance of continuous long-term evolution of the coronal field, rather than independent extrapolations at different times.

There are several improvements that we hope to make in future simulations. Firstly, in view of the results above, future simulations will have a longer ``ramp-up'' time from the initial potential field. Further improvements concern the observations of newly-emerging active regions, which are crucial to our model. In the present study, all newly-inserted regions were given the same magnitude of twist in any individual simulation run. However, using photospheric vector magnetograms, we hope in future to measure the twist of individual active regions \cite{nandy2007} and incorporate this into the simulations. Another uncertainty in our model is the exact date of emergence of each active region, but this can not be seriously addressed without observations of the far-side of the Sun, although helioseismic imaging techniques \cite{lindsey2000} offer some promise in this respect. 

In summary, this study has demonstrated the importance of long-term evolution and build-up of non-potential magnetic structure in the solar corona. The transport of helicity from low to high latitudes over many months is a fundamental element in the coronal evolution. After the successful reproduction of filament chirality, we hope that the model will be useful for further study of the corona. An interesting feature to explore is the temporary loss of equilibrium and ejection of magnetic flux ropes, as reported by \inlinecite{mackay4}; such behaviour also occurs in our global simulations. These eruptions may correspond to coronal mass ejections \cite{zhang2005,hudson2006}, and an understanding of the magnetic topologies responsible offers the possibility of space weather forecasting on the time-scale of weeks or months.

\begin{acks}
Financial support for ARY and DHM was provided by the UK Science and Technology Facilities Council (STFC). The simulations were performed on the UKMHD parallel computer in St Andrews, funded jointly by SRIF/STFC. We would like to thank S.F. Martin for supplying the original filament data set, based on data from BBSO/NJIT, as well as travel support from the PROM group (NSF grant ATM-0519249). Synoptic magnetogram data from NSO/Kitt Peak was produced cooperatively by NSF/NOAO, NASA/GSFC, and NOAA/SEL and made publicly accessible on the World Wide Web. Finally we thank the anonymous referee for his/her detailed comments which have improved the paper.
\end{acks}

\end{article} 
\end{document}